\documentclass[structabstract]{aa}  
\usepackage{graphicx}
\usepackage{times}
\usepackage{epsfig}
\usepackage{longtable}
\usepackage{lscape}
\usepackage{amssymb}
\usepackage{latexsym}
\usepackage{natbib}
\usepackage{txfonts}
\def\kms{$\rm km\;s^{-1}$}

\def\hb{H$\beta$}

\def\mgb{Mg~{\it b}}
\def\fei{Fe{\small 5270}}
\def\feii{Fe{\small 5335}}
\def\feiii{Fe{\small 5105}}
\def\feiv{Fe{\small 5406}}
\def\fev{Fe{\small 5709}}
\begin{document}
\title{Stellar population and the origin of intra-cluster stars around
  brightest cluster galaxies: the case of NGC 3311\thanks{Based on
    observations collected at the European Southern Observatory for
    the program 082.A-0255(A).}}

   \subtitle{}

   \author{L. Coccato \inst{1}
          \and O. Gerhard \inst{2}
	  \and M. Arnaboldi \inst{1,3}
          \and G. Ventimiglia \inst{2}
          }

   \offprints{L. Coccato, e-mail: lcoccato@eso.org}

   \institute{European Southern Observatory, Karl-Schwarzschild-Stra$\beta$e 2,
          85748 Garching bei M\"unchen, Germany.
    \and Max-Plank-Institut f\"ur Extraterrestrische Physik,
     Giessenbachstra$\beta$e 1, D-85741 Garching bei M\"unchen, Germany.
    \and INAF, Osservatorio Astronomico di Pino Torinese, I-10025 Pino Torinese, Italy.
}
   \date{\today}

 
  \abstract
{We investigate the stellar population and the origin of diffuse light around brightest cluster galaxies.}
{We study the stellar population of the dynamically hot stellar
halo of NGC 3311, the brightest galaxy in the Hydra I cluster, and
that of photometric substructures in the diffuse light to
constrain the origin of these components.}
 {We analyze absorption lines in medium-resolution, long-slit spectra
  in the wavelength range 4800 -- 5800 \AA\ obtained with FORS2 at the
  Very Large Telescope. We measure the equivalent width of Lick
  indices out to 20 kpc from the center of NGC~3311 and fit them with
  stellar population models that account for the [$\alpha$/Fe]
  overabundance.}
{Stars in the dynamically hot halo of NGC 3311 are old (age $> 13$
  Gyr), metal-poor ([Z/H] $\sim-0.35$), and alpha-enhanced
  ([$\alpha$/Fe] $\sim 0.48$). Together with the high velocity
  dispersion, these measurements indicate that the stars in the halo
  were accreted from the outskirts of other early-type galaxies, with
  a possible contribution from dwarf galaxies. We identify a region in
  the halo of NGC 3311 associated with a photometric substructure
  where the stellar population is even more metal-poor
  ([Z/H]$\sim-0.73$). In this region, our measurements are consistent
  with a composite stellar population superposed along the line of
  sight, consisting of stars from the dynamically hot halo of NGC 3311
  and stars stripped from dwarf galaxies. The latter component
  contributes $\leq 28$\% to the local surface brightness.}
{ The build-up of diffuse light around NGC 3311 is on-going. Based
    on the observed stellar population properties, the dominant part
    of these stars may have come from the outskirts of bright
    early-type galaxies, while stars from stripped dwarf galaxies are
    presently being added.}


   \titlerunning{Stellar populations in BCG halos: NGC 3311}

   \authorrunning{Coccato et al.}

   \maketitle
%

\section{Introduction}

Brightest cluster galaxies (BCGs) are the most massive and luminous
objects located at the center of galaxy clusters. According to the
hierarchical formation scenario, they assemble through the merger of
smaller objects that are dragged into the cluster's center by dynamical
friction (e.g. \citealt{DeLucia+07}). During the evolution of the
cluster and the formation of the central BCG, stars are stripped from
their host galaxies. Part of the stripped material is captured by the
halos of the BCG and other massive galaxies, while the other part
remains free-floating in the cluster potential, forming the so-called
intra-cluster light (ICL). The details of the interplay between
the formation of the BCG halos and the ICL are not yet fully understood.
Observations of both components can give key information about the
processes involved in the evolution of cluster galaxies.

In the last few years, progress was achieved in the characterization
of the stellar population properties of the halos of BCGs at
galactocentric radii larger than three effective radii by means of
long-slit spectroscopy \citep{Brough+07, Coccato+10a, Coccato+10b} or
photometric studies of globular clusters \citep{Forbes+11,
  Arnold+11}. Stars in the BCG halos are generally old ($\geq$10 Gyr),
they have solar or sub-solar metallicity and super-solar
$\alpha-$enhancement. Sometimes a change in the radial profiles of
metallicity, $\alpha-$enhancement or color is observed, indicating
that the stellar population properties in the outskirts are different
from those in the central regions. The emerging interpretation is that
the inner and outer regions formed in two separate phases: the
central parts formed ``in situ'' (or through a single major merger),
while the halo was accreted later through a series of minor mergers.
This ``two stages'' formation process for BCGs is also supported by
numerical simulations (e.g. \citealt{Abadi+06, Oser+10}) and
photometric studies that compare the BCGs scaling relations between
color gradients, velocity dispersions, and sizes with those of other
early-type galaxies of similar mass or luminosity \citep{Bernardi+11}.
Minor merger accretion in the second stage of this scenario is also
consistent with the observed size evolution of early-type galaxies
over redshift \citep{Trujillo+07, vanDokkum+10}, hydrodynamical
cosmological simulations \citep{Naab+09}, and spectroscopic studies
(\citealt{Coccato+10b}; \citealt{Trujillo+11}).

It is not clear yet whether the major contribution for the formation
of the ICL comes from stars stripped in galaxy groups and subsequently
captured by the cluster potential \citep{Rudick+06, Rudick+09,
  Kapferer+10}, from stars of galaxies that interact with each other
and with the cluster tidal field \citep{Moore+96, Willman+04}, from
stars stripped from galaxies merging into the central BCG
\citep{Murante+07, Puchwein+10}, or from entirely disrupted low
surface brightness galaxies or dwarfs \citep{Gnedin03}.

\begin{figure}
 \psfig{file=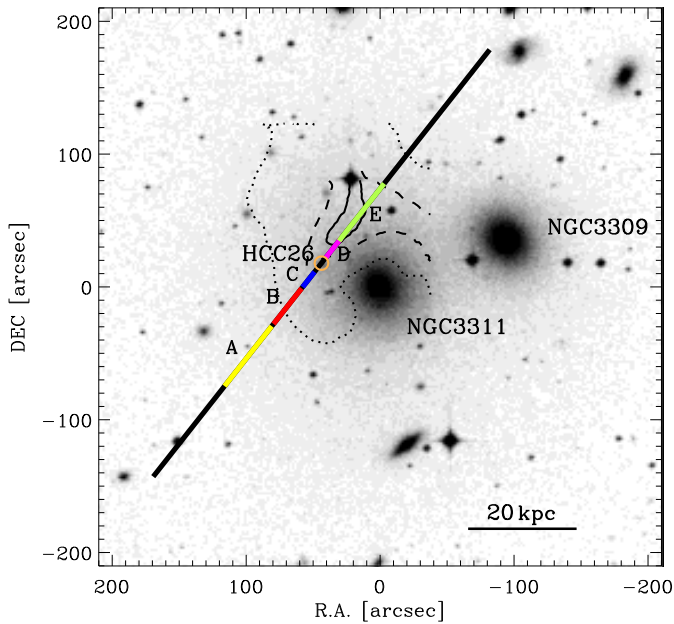,width=8.8cm,clip=}
\caption{Digital Sky Survey image of the central regions of the Hydra
  I cluster. The two brightest galaxies NGC~3311 and NGC~3309 are
  labeled. The thick black line represents the location of the FORS2
  long slit. The orange circle marks the slit center on top of the
  dwarf galaxy HCC~26. The spectrum of HCC~26 was extracted from the
  central 6'' (see text for details). The colored lines identify the
  location of the five slit sections (bins A-E) where we extracted the
  spectra of the halo of NGC3311.  Thin continuous, dashed and dotted
  lines represent approximately the contours of a photometric
    substructure found by \citet{Arnaboldi+11}. We adopt a distance
  to NGC 3311 of 51 Mpc, from the NASA/IPAC extragalactic database.}
\label{fig:slit_location}
\end{figure}

The connection between the accretion of the BCG halo and the formation
of the ICL is hard to probe from an observational point of view, given
the low surface brightness of these regions. \citet{Williams+07}
measured the stellar population parameters for a sample of $\sim$5300
intra-cluster stars in one field in the Virgo cluster. Although they
found a heterogeneous stellar population, 70\%--80\% of the observed
intra-cluster stars are old (age $> 10$ Gyr) and metal-poor (mean
[M/H]$ \sim -1.0$).  On the other hand, other studies report super-solar
metallicities in the diffuse light in Abell 3888 ($1.0 Z_{\odot} < Z <
2.5 Z_{\odot}$, \citealt{Krick+06}) that are interpreted as ICL formed
early with the collapse of the main cluster, unrelated to the BCG
halos.  From a kinematic point of view, the line-of-sight velocity
distribution of intra-cluster planetary nebulae in one Coma cluster
core field is consistent with the ICL being formed from stars stripped
from the halos of the two brightest galaxy in the cluster while they
orbit each other \citep{Gerhard+07}, while recent studies of
absorption line spectra in the RX J0054.0-2823 cluster are consistent
with the ICL being the remnant of tidally destroyed galaxies and
streaming in the central regions of the cluster, in which the three
central giant ellipticals act as "grinding machines"
\citep{Toledo+11}. The origin of the ICL and its connection with the
BCG halos is thus not fully understood, and it is likely that several
mechanisms are acting in its formation.

In this context we have been carrying out a project aimed at studying
the formation of the intra-cluster stellar component in the Hydra I
galaxy cluster (Abell 1060), and this paper is the fourth in this
series. We observed a rapid increase of the velocity dispersion radial
profile in the halo of NGC~3311, the BCG in Hydra I
\citep{Ventimiglia+10}. This was interpreted as caused by
a dynamically hot stellar halo around NGC 3311, dominated by
intra-cluster stars.  We also observed the velocities of planetary
nebulae in the core of the Hydra I cluster, the major component of
which is consistent with being part of this halo of intra-cluster
stars \citep{Ventimiglia+11}.  The distribution and kinematics of the
entire sample of intra-cluster planetary nebulae indicate that most of
the diffuse light is still not phase-mixed. Finally,
\citet{Arnaboldi+11} found that the hot stellar halo is not centered
on NGC 3311, and detected a photometric sub-structure (excess of
light, hereafter) superposed on the intracluster halo. The contours of
light that best approximate this photometric excess are shown in
Figure \ref{fig:slit_location}. Their results indicate that this
excess of light has line-of-sight velocity $\sim$ 1200 \kms\ relative
to the stellar halo around NGC~3311, and that it is related to one or
more dwarf galaxies disrupted in this region. The build-up of the ICL
and the dynamically hot stellar halo around NGC~3311 are therefore
ongoing through the accretion of material from  galaxies falling into
the cluster core and tidally interacting with its potential well.

In this paper we investigate the nature of the dynamically
hot stellar halo around NGC~3311 and the excess of light by
studying their stellar populations. Section \ref{sec:observations}
presents the spectroscopic observations and the data reduction
steps. Section \ref{sec:population} describes measurements of
equivalent widths of the Lick indices and the derivation of stellar
age, metallicity, and $\alpha$-enhancement. In Section
\ref{sec:discussion}, we discuss the origins of the stellar halo of NGC
3311 and the excess of light, and finally in Section \ref{sec:summary}
we conclude with a brief summary.

\begin{figure*}
\vbox{
 \hbox{
   \psfig{file=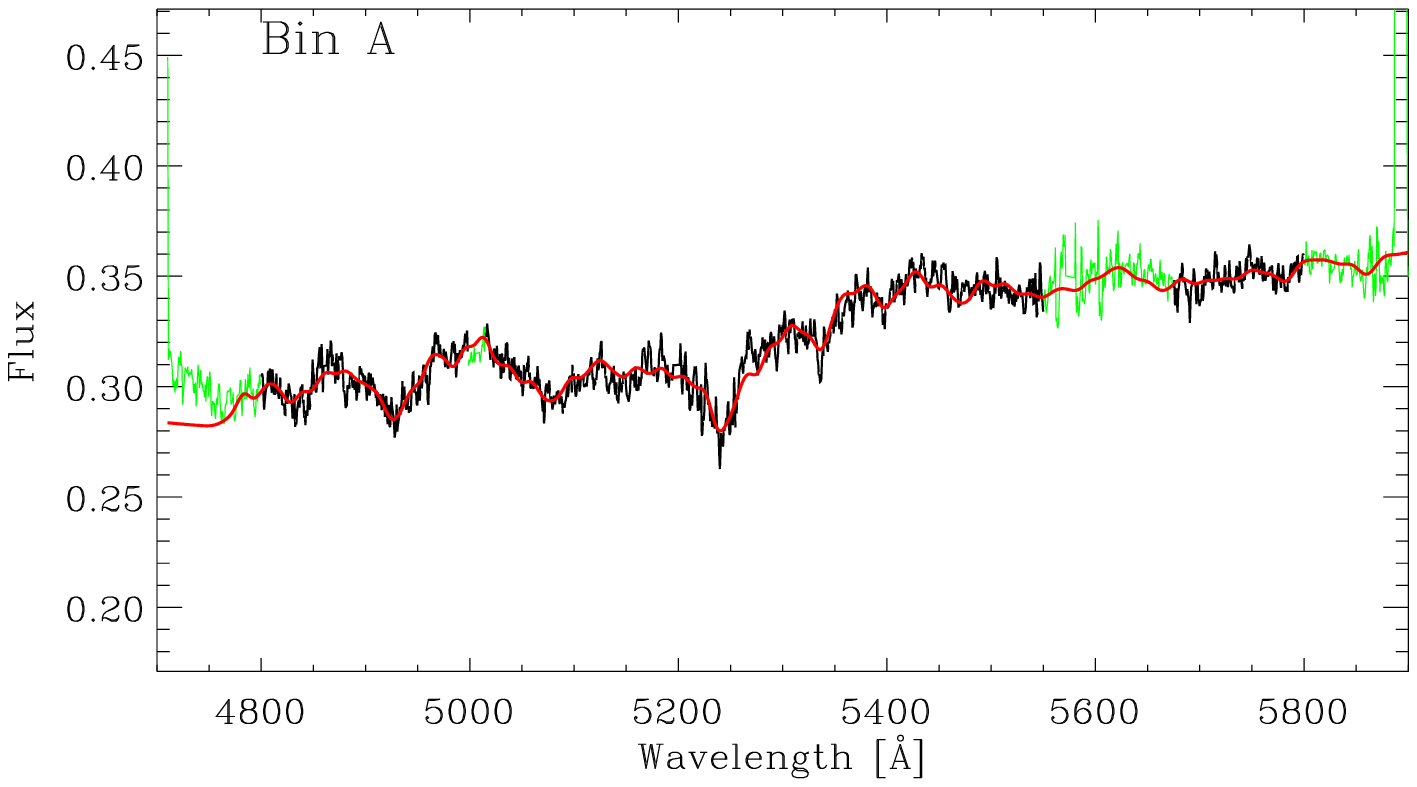,width=9.0cm,clip=}
   \psfig{file=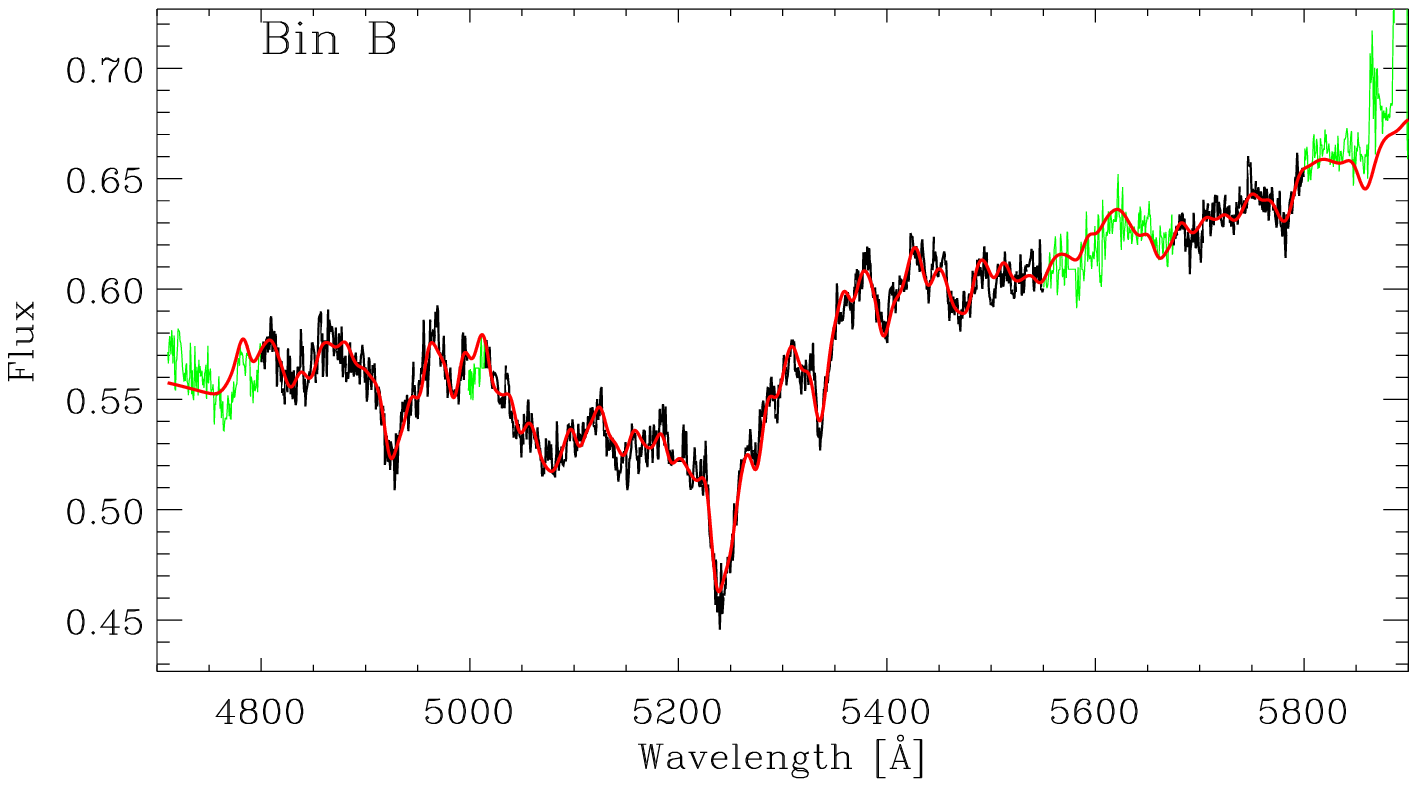,width=9.0cm,clip=}}
 \hbox{
   \psfig{file=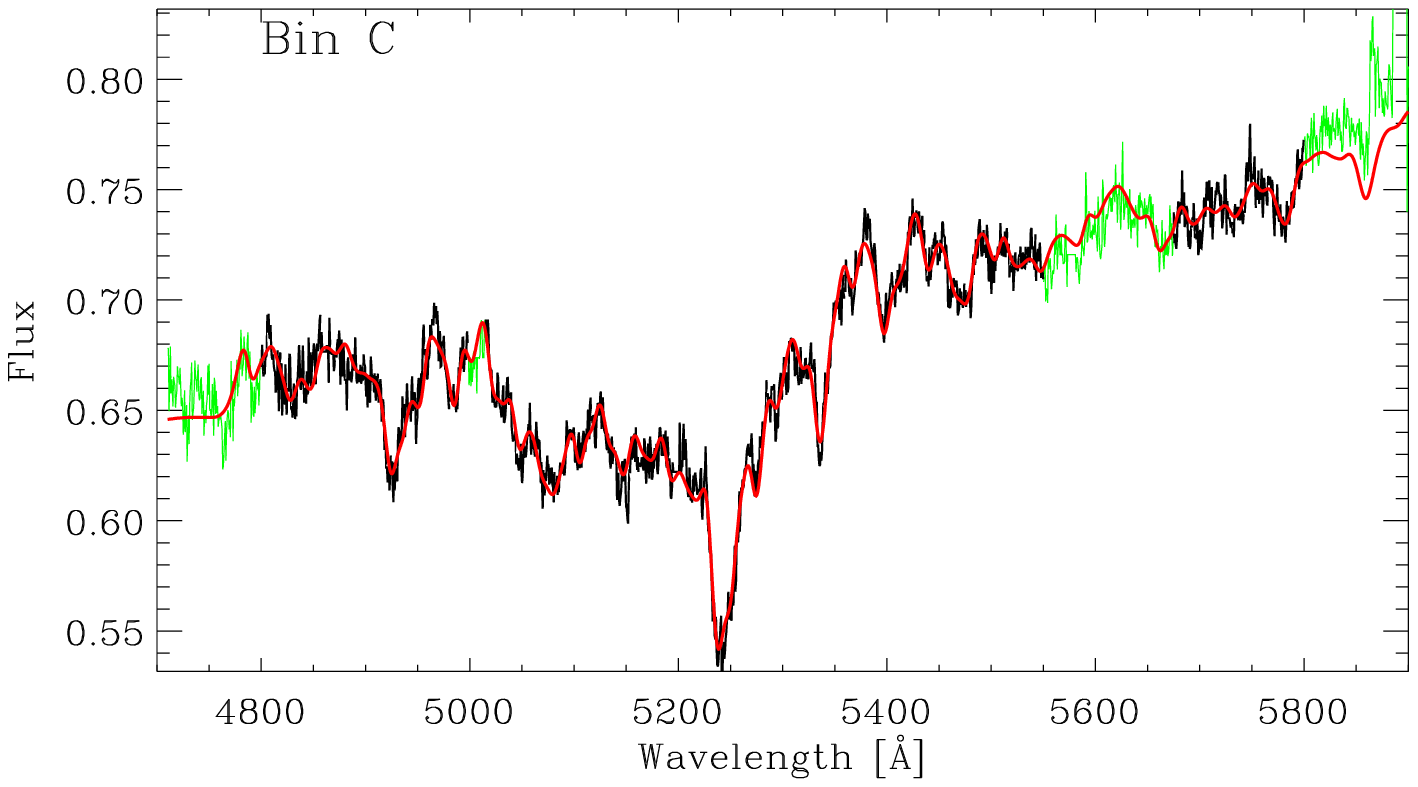,width=9.0cm,clip=}
   \psfig{file=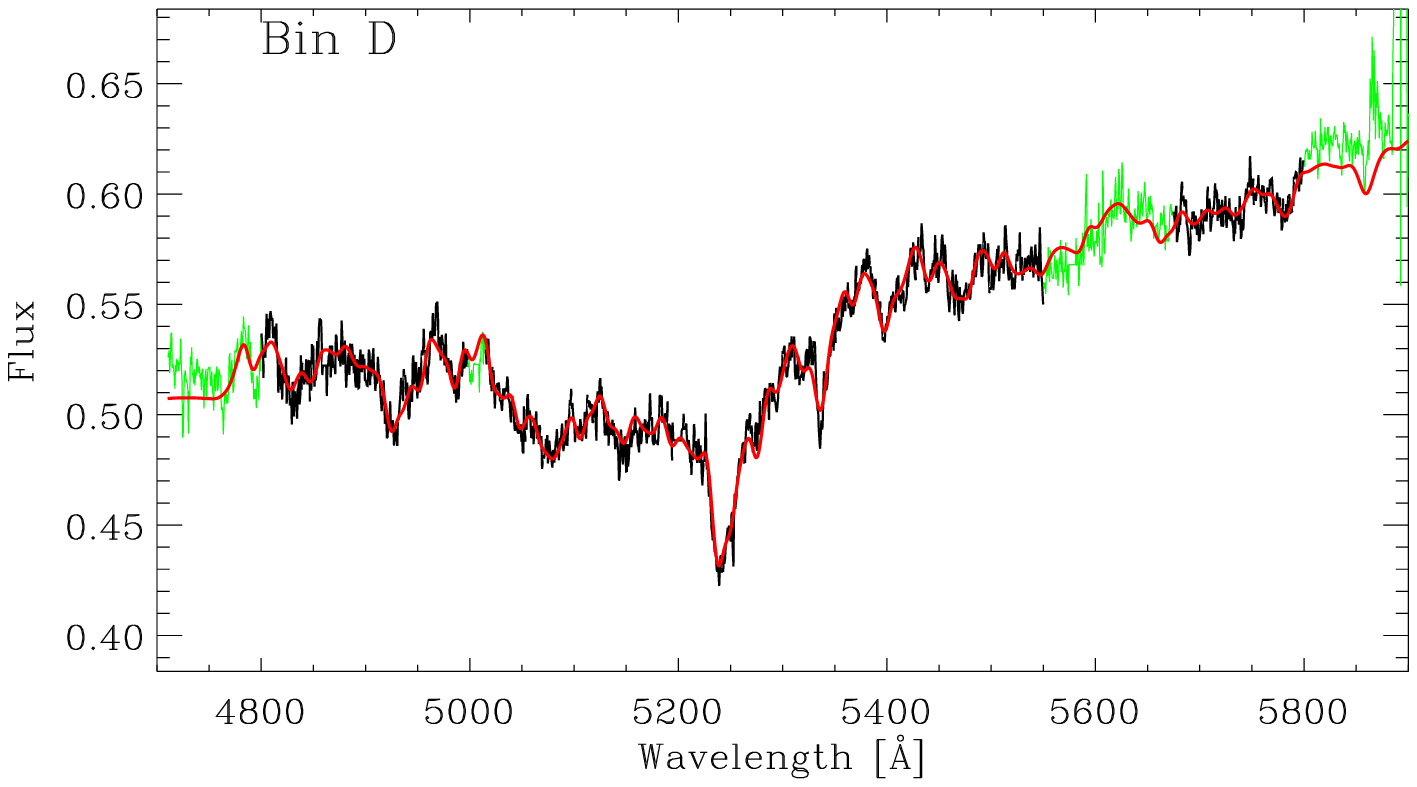,width=9.0cm,clip=}}
\hbox{ 
\psfig{file=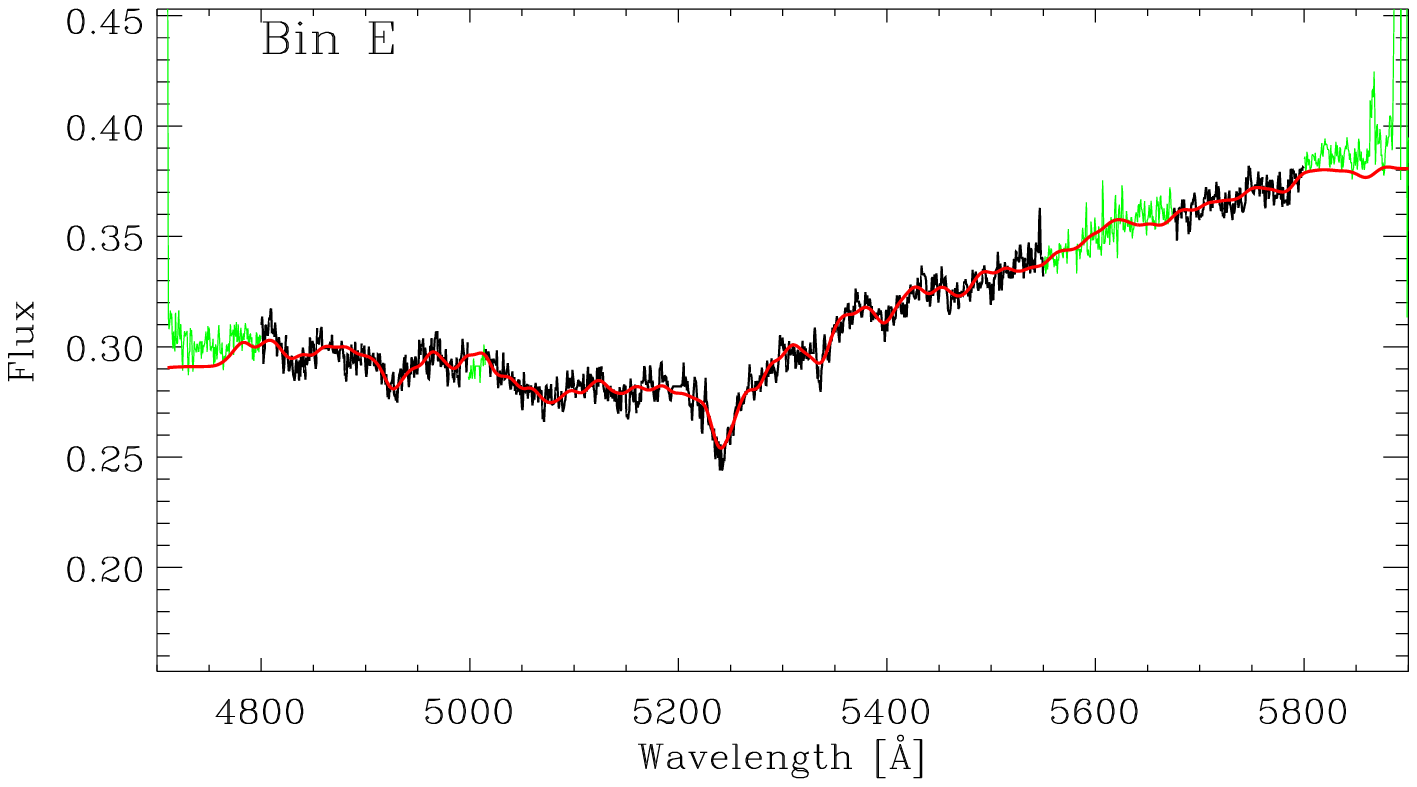,width=9.0cm,clip=}
\psfig{file=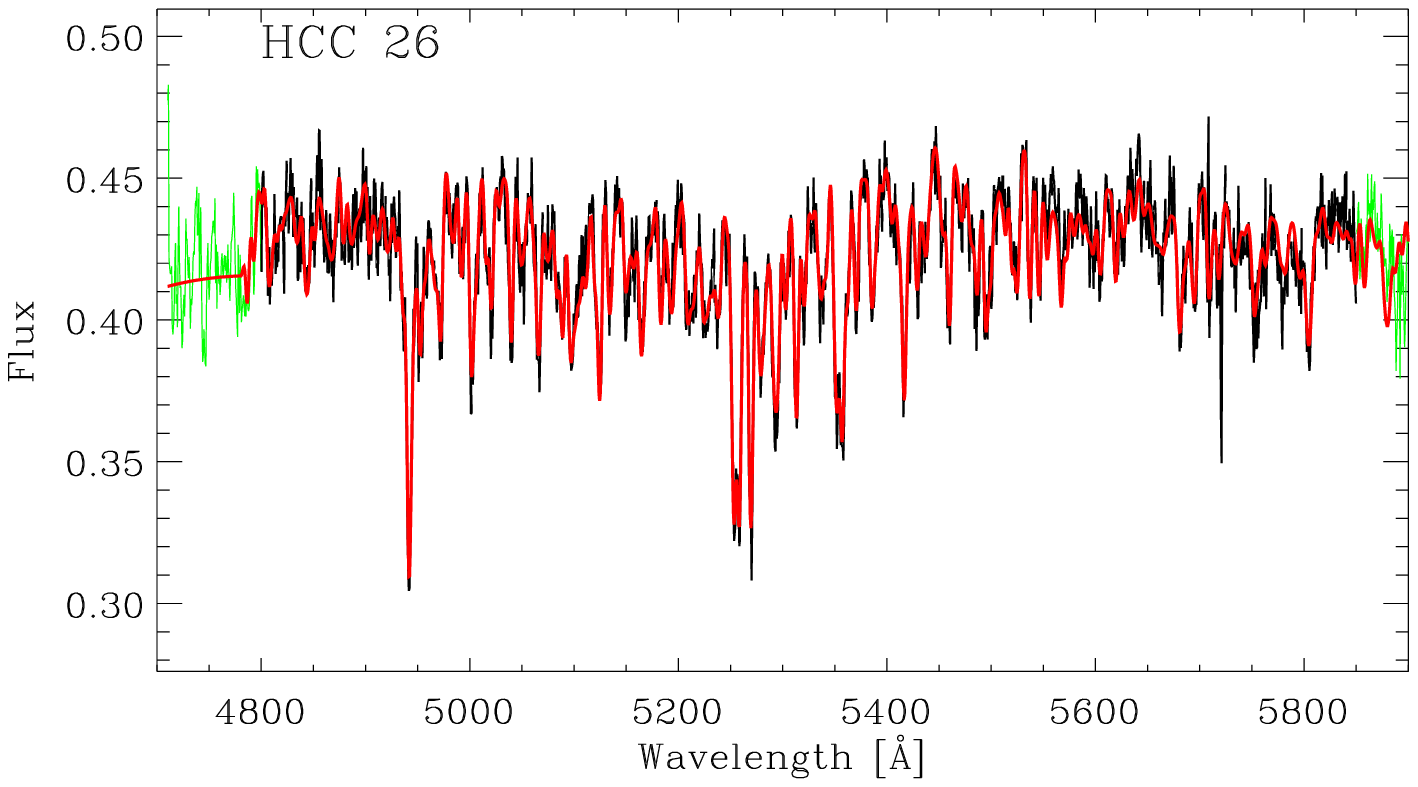,width=9.0cm,clip=}}
}
\caption{Spectra extracted in the selected slit sections (bins A-E)
  and in the slit center (HCC~26). {\it Black:} observed spectrum,
  {\it red:} best-fit model, {\it green} wavelength region excluded
  from the fit.}
\label{fig:fit}
\end{figure*}

\section{Spectroscopic observations and data reduction}
\label{sec:observations}

The long-slit spectra were obtained during the nights of March
25--28--2009, with the focal reducer and low-dispersion spectrograph
(FORS2, \citealt{Appenzeller+98}) mounted on the Unit Telescope 1 of
the European Southern Observatory (ESO) in Paranal (Chile). The
long-slit was oriented at position angle $PA=142^{\circ}$, and
centered on the dwarf galaxy HCC~26, which is $\sim 45''$ northeast
from NGC~3311 (Fig. \ref{fig:slit_location}). The instrumental setup
had the $1\farcs6$ wide long-slit and the 1400V+18 grism with
instrumental spectral resolution of $FWHM \sim 210$ \kms. The total
exposure time was four hours. For more information on the instrumental
set-up see \citet{Ventimiglia+10}.

Data reduction (overscan and bias subtraction, flat-fielding
correction, wavelength and flux calibrations) was performed with the
FORS2 reduction pipeline version 3.0 under the ESOrex
environment\footnote{Updated version of the FORS2 pipeline and ESOrex
  can be found at http://www.eso.org/sci/software/pipelines/} and {\tt
  IRAF}\footnote{{\tt IRAF} is distributed by NOAO, which is operated
  by AURA Inc., under contract with the National Science Foundation.}
and {\tt IDL} {\it ad hoc} scripts.  The contribution from scattered
light in the instrument was interpolated from the portions of the
FORS2 detectors not illuminated by the long-slit (typically a few rows
on the CCDs borders) and then subtracted. Scattered light amounted to
$7 \pm 3$ \% of the sky-background value, with negligible gradients
along the dispersion direction.  Wavelength calibration and spectral
rectification were performed using He, Hg+Cd, and Ne calibration lamp
spectra and were cross-checked with the most intense emission lines in the
sky spectrum. Errors on the wavelength calibration were $\sim 5$
\kms. Cosmic rays were removed using the {\tt FILTER/COSMIC} routine
in {\tt MIDAS}\footnote{{\tt MIDAS} is developed and maintained by
  ESO.}.  Long-slit spectra on blank fields in the Hydra I cluster
were also obtained close in time to the scientific spectra to evaluate
the sky background and to correct for large-scale illumination
patterns from the non-uniform illumination along the slit.  All
spectra were corrected for atmospheric extinction and flux-calibrated
using spectro-photometric standard stars observed during twilight as
part of the FORS2 standard calibration plan.

\begin{table*}
\centering
\caption{Line strengths of Lick indices in the hot stellar halo around NGC 3311.}
\begin{tabular}{l c c  c  c  c  c  c c  c}
\hline \noalign{\smallskip}
   ID      & $D$ &	$PA$    & \hb	      &Fe{\small 5015}& \mgb          & \fei          & \feii         &Fe{\small 5406}&Fe{\small 5709}\\
           & [$''$]& [$^{\circ}$]& [\AA]       & [\AA]         & [\AA]         &  [\AA]        &  [\AA]	  & [\AA]	  & [\AA]         \\
   (1)     &  (2)  &     (3)    &    (4)      &     (5)       &   (6)         &     (7)       &   (8)        &   (9)        &  (10)          \\   
\hline 
\noalign{\smallskip}
   D     &42.8  &     47.1   &$0.77\pm0.09$ &	$3.14\pm0.19$ &	$3.02\pm0.12$ &	$1.63\pm0.10$ &	$1.84\pm0.17$ &	$1.13\pm0.09$ &	$0.81\pm0.08$ \\
   C     &50.2  &     83.7   &$0.96\pm0.09$ &	$3.42\pm0.18$ &	$3.41\pm0.10$ &	$1.69\pm0.10$ &	$2.05\pm0.15$ &	$1.26\pm0.09$ &	$0.85\pm0.08$ \\
   E     &57.1  &     10.3   &$0.66\pm0.10$ &	$2.57\pm0.22$ &	$2.57\pm0.18$ &	$1.24\pm0.11$ &	$1.58\pm0.21$ &	$0.77\pm0.12$ &	$0.75\pm0.10$ \\
   B     &67.2  &    102.5   &$1.11\pm0.09$ &	$3.40\pm0.19$ &	$3.18\pm0.11$ &	$1.78\pm0.10$ &	$1.80\pm0.15$ &	$1.23\pm0.09$ &	$0.81\pm0.08$ \\
   A     &108.4 &    118.8   &$0.99\pm0.10$ &	$4.05\pm0.24$ &	$3.06\pm0.26$ &	$1.68\pm0.18$ &	$1.61\pm0.23$ &	$0.77\pm0.13$ &	$0.69\pm0.11$ \\
 \noalign{\smallskip}                   
 \noalign{\smallskip}                   
HCC26   &  45.0 &     64.0   &$1.85\pm0.09$ &	$2.53\pm0.18$ &	$1.85\pm0.07$ &	$2.05\pm0.07$ &	$1.87\pm0.13$ &	$1.22\pm0.07$ &	$0.61\pm0.09$ \\
 \noalign{\smallskip}                   
\hline                                 
\noalign{\smallskip}
\end{tabular}                          
\label{tab:indices}                 
 \begin{minipage}{18cm}                 
  Notes -- Col. 1: Slit bin name, according to Fig. \ref{fig:slit_location}.
           Col. 2: Distance of the bin center from the center of 
                   NGC 3311
           Col. 3: Position angle of the  bin center with respect to 
                   the center of NGC 3311. North is $0^{\circ}$, East is $90^{\circ}$.
           Cols.4-11: Equivalent widths of the Lick indices and their errors.
\end{minipage}                         
\end{table*}

\subsection{Sky subtraction}
\label{sec:sky}

Accurate sky subtraction is important when measuring spectra of low
surface brightness regions. To limit systematic effects as much as
possible, we accounted for the sky-subtraction in two ways, following
the strategy adopted in \citet{Coccato+10a}.  The mean flux of the
continuum of the sky spectrum extracted from the outermost portions of
the long slit was $\sim 25\%$ higher than that of the sky spectrum
extracted from blank fields in the Hydra I cluster.  This indicated
that the contributions from the halo of NGC~3311 and from
intra-cluster stars were not negligible in the long-slit ends. We
therefore decided to use the sky spectrum obtained from blank
fields. Although they were taken close in time to the scientific
spectra, they were not simultaneous. To minimize the effects of the
variation in time of the intensity of the sky emission lines, we
proceeded in a way similar to that of \citet{Davies+07}.  For each
night's sky spectrum, a fourth order polynomial was fitted to the
continuum and then subtracted. The mean flux of the sky-background
continuum in the $4800 - 5600$ \AA\ range did not vary by more than
$\pm 5\%$ from night to night. The residual spectrum that contained
emission and absorption line features was multiplied by a constant
factor and added to the sky continuum previously subtracted. The
constant factor (typically on the order of $1.20 - 1.30$) was chosen
 to minimize the difference between the intensities of some
selected sky emission lines measured in the scientific spectra and
those measured in the sky spectra. The entire process was optimized in
the wavelength range 4800 \AA\ -- 5800 \AA.

Finally, the sky-subtracted spectra were coadded using the center of
HCC~26 as reference for alignment.

\smallskip

{\it HCC~26.} The spectrum of HCC~26 was extracted from the central
$6''$ of the FORS2 long-slit (orange circle in
Fig. \ref{fig:slit_location}). The contribution of NGC 3311 was
interpolated from the adjacent $\sim$5\farcs8, where residual flux of
HCC26 is negligible (lower than $4\%$ of the HCC~26 central value and
comparable to the local noise level) and removed from the spectrum.

\smallskip 

{\it NGC 3311 halo.} The spectrum of the halo of NGC 3311 was
extracted along five sections of the FORS2 long-slit (bins A-E) from the
sky-subtracted spectrum. Their locations are displayed in
Fig. \ref{fig:slit_location}. The edges of the central bins C and D
are $\sim 6''$ away from the slit center, at positions where the
contamination from HCC~26 is lower than the background noise level.
The sizes of the bins are chosen to ensure a minimal
signal-to-noise ratio $\sim 25$ per spectral pixel in each bin, and to
account for the CCD bad cosmetics and faint foreground stars in the
slit. Outside the explored region the surface brightness level is too
small to permit reliable measurements.

\medskip

We quantified the residual contamination from sky emission lines 
  on the final co-added science spectra to be lower than $2\%$ of the
subtracted sky itself, and lower than the noise level. This was
determined in the following way:
 
\begin{enumerate}

\item We subtracted the best-fit models from the observed galaxy
  spectra, obtaining the residual spectra $R(\lambda)$ for each slit
  bin. We computed the average noise $<{\rm rms}>$ values as standard
  deviations of the $R(\lambda)$ spectra in the wavelength range 4900
  -- 5550 \AA.

\item We cross-correlated the $R(\lambda)$ with the
  continuum-subtracted sky spectrum $S_{\rm ems}(\lambda)$ to detect
  evidence of contamination from residual emission lines. No
  significant correlation was found, meaning that the contamination
  was below the noise level.

\item We constructed a noise spectrum $N(\lambda)$ with mean 0 and
  standard deviation equal to the maximum $<{\rm rms}>$, which is
  measured at the outermost slit section (bin A, Fig. \ref{fig:fit}).

\item We determined the minimum value of $K$ that ensured a
  significant correlation between the $S_{\rm ems}(\lambda)$ spectra
  and $N(\lambda) + K\cdot S_{\rm ems}(\lambda)$. We found $K=0.02$,
  meaning that the highest possible contamination from residual
    sky-emission lines is $2\%$ of the sky value. Indeed, if the
  contamination were higher, it would have been detected in step
  2. The 2\% upper limit will be used in Section \ref{sec:ssp} to
  quantify the maximum systematic errors caused by residual sky
  contamination in the derived stellar population parameters.

\end{enumerate}

\section{Line strength indices and single stellar population}
\label{sec:population}
The spectra extracted  from the slit sections were fitted
with the MILES library of stellar template spectra \citep{Sanchez+06}
using the penalized pixel-fitting routine by \citet{Cappellari+04} to
recover the stellar velocity ($V$), velocity dispersion
($\sigma$), third and fourth Gauss-Hermite moments ($h3$ and $h4$) of
  the line-of-sight velocity distribution (LOSVD). In the fitting
process, the spectra of the MILES library were convolved with a
Gauss-Hermite function to match the instrumental FORS2 spectral
resolution. The appropriate function was determined by fitting
Gauss-Hermite moments to the correlation function between the spectra
of one star observed during the run and its corresponding spectrum in
the MILES library. The use of a Gauss-Hermite function instead
  of the commonly used Gaussian function ensured more reliable
measurements, because the FORS2 instrumental line-spread function
slightly deviates from a Gaussian profile. The extracted spectra
are shown in Figure \ref{fig:fit} together with their best-fit models.
We refer the reader to \citet{Ventimiglia+10} for the discussion of
the kinematics. In this paper we focus on the measurements of the Lick
indices and the derivation of the stellar population properties.

Galaxy spectra were set to the rest wavelength using the measured
radial velocities and then broadened to match the Lick instrumental
spectral resolution (8.4 \AA\ at 5100 \AA, \citealt{Worthey+97}). The
equivalent widths of the line strengths of several Lick indices were
measured according to the definitions of \citet{Worthey+94}, and their
errors were evaluated using the empirical formulas by
\citet{Cardiel+98}.

The measurements were also corrected for kinematic broadening caused
by the line-of-sight velocity distribution, using the multiplicative
coefficients tabulated in \citet{Kuntschner04}, which account for
$\sigma$, $h3$ and $h4$. Because no coefficient is tabulated for the
\fev, the corresponding factor was determined as the ratio of i) the
“intrinsic” values measured on the optimal stellar template; and ii)
the values measured on the optimal stellar template convolved with the
measured line of sight velocity distribution.

A set of seven Lick standard stars were also observed with the same
instrumental configuration. The line strengths of the Lick indices
were measured for these stars and compared with the values tabulated
in \citet{Worthey+94} to measure offsets to the Lick system.
Small offsets were observed for the \hb\ ($-0.11 \pm 0.08$ \AA)
and \mgb\ ($-0.13 \pm 0.07$ \AA) indices and were applied to our
measurements.

Table \ref{tab:indices} lists the final equivalent widths of the Lick
indices we measured, corrected for kinematic broadening and offset to
the Lick system. Errors in the equivalent widths also include the
uncertainties in the offsets to the Lick system and in the 
  correction for kinematic broadening.

\subsection{Luminosity-weighted single stellar population}
\label{sec:ssp}

We compared our measured Lick indices \hb, \mgb, \fei, \feii, \feiii,
\feiv, and \fev\ to the values predicted by the single stellar
population models of \citet{Thomas+11}. We interpolated the original
grid of models so that we have steps of $\sim 0.4$ Gyr in age and
$\sim 0.025$ dex in metallicity ([Z/H]) and in $\alpha$-enhancement
([$\alpha$/Fe]). The best stellar population parameters are obtained
by a $\chi^2$ minimization using the observed indices.  Errors on the
best-fit parameters are computed by means of Monte Carlo simulations.

To quantify the errors on age, [Z/H] and [$\alpha$/Fe] caused by
residual sky contamination in the spectra (see Sect. \ref{sec:sky}) we
added to the scientific spectra an extra contribution $\pm 2\%$ of the
subtracted sky and measured the single stellar population parameters
again. The best-fit luminosity weighted values of age, [Z/H] and
[$\alpha$/Fe] and the range allowed by potential contamination from
sky emission lines are shown in Figure \ref{fig:ssp} and listed in
Table \ref{tab:ssp}.

There are no data in the literature that extend from the center of NGC
3311 toward the halo and overlap with our measurements, but according
to \citet{Loubser+08}, the central region of NGC 3311 is younger
(Age$_0$$=8.7\pm1.8$ Gyr) considerably more metal-rich
([Z/H]$_0$$=0.12\pm0.07$) than what we observe in the hot halo, and
has similar abundance ratio ([$\alpha$/Fe]$_0$$=0.40\pm0.03$).

\begin{table}
\centering
\caption{Luminosity-weighted single stellar population
  parameters in the hot stellar halo around NGC 3311.}
\begin{tabular}{c c  c  c c}
\hline \noalign{\smallskip}
 ID&$a$   &  Age                &    [Z/H]        & [$\alpha$/Fe]  \\ 
   &[$''$]&  [Gyr]              &                 &                \\ 
 (1)&   (2)&   (3)               &       (4)       &  (5)           \\  
\hline 
\noalign{\smallskip}
D & 43.2  & $13.5^{+0.0}_{-0.8}$ & $-0.46 \pm 0.04$& $0.45 \pm 0.04 $\\
C & 50.8  & $13.5^{+0.0}_{-0.8}$ & $-0.33 \pm 0.03$& $0.48 \pm 0.03 $\\
E & 61.3  & $13.5^{+0.0}_{-0.8}$ & $-0.73 \pm 0.06$& $0.50 \pm 0.06 $\\
B & 70.2  & $13.5^{+0.0}_{-0.8}$ & $-0.39 \pm 0.04$& $0.44 \pm 0.04 $\\
A &117.0  & $13.5^{+0.0}_{-0.8}$ & $-0.34 \pm 0.05$& $0.50 \pm 0.03 $\\
 \noalign{\smallskip}                   
 \noalign{\smallskip}                   
HCC~26&45.4& $13.5^{+0.0}_{-1.0}$ & $-0.85 \pm 0.03$& $-0.03 \pm 0.05$\\
\noalign{\smallskip}                   
\hline                                 
\noalign{\smallskip}
\end{tabular}                          
\label{tab:ssp}                 
 \begin{minipage}{9cm}               
  Notes -- Col. 1: bin name, according to Fig. \ref{fig:slit_location}.
        Col. 2: distance of the bin center from the center of NGC
                3311 projected onto the galaxy major axis assuming
                an axial ratio $b/a=0.89$ \citep{Arnaboldi+11}.
        Cols. 3-4: single stellar population best-fit values and their 
                   errors. [Z/H] and [$\alpha$/Fe] are in logarithm of 
                   solar units.
\end{minipage}                         
\end{table}

\subsection{Contamination from foreground stars}

\begin{figure}
 \psfig{file=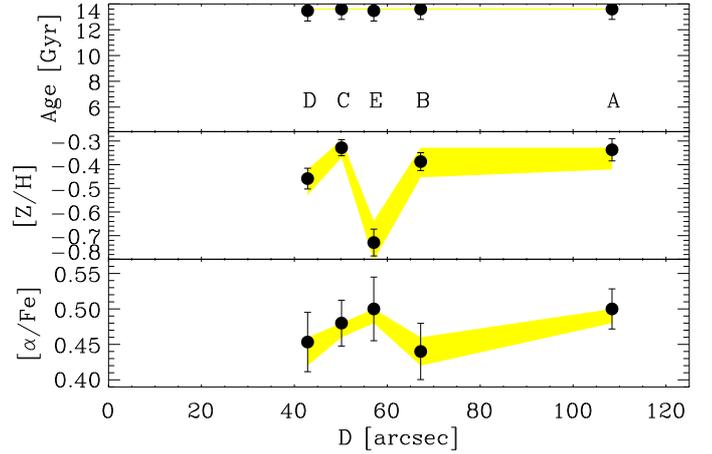,width=9.0cm,clip=}
\caption{Equivalent luminosity-weighted single stellar population
  values of age (upper panel), metallicity (central panel) and
  $\alpha$-enhancement (lower panel) obtained in the five long-slit
  sections, as function of the distance to NGC 3311. Labels in the upper
  panel identify each radial bin. The shaded yellow regions indicate
  the range of SSP parameters that would result if systematic residual
  effects in the sky subtraction at the level of $\leq \pm 2$\% were
  present (see Sect. \ref{sec:ssp} for details).}
\label{fig:ssp}
\end{figure}

At $\sim30''$ north of the center of bin E there is a bright star
with $B \sim 12.2$ mag from the USNO-B1 catalog but unknown spectral
classification. Although this star does not enter the slit directly 
(its distance to the long-slit is $\sim$20 times the seeing FWHM), it
is inside the FORS2 field of view and its light can be scattered along
the optical path of the spectrograph and hit the detector. In
Sect. \ref{sec:observations} we quantified the amount of scattered
light ($\sim 7 \pm 3$\% of the sky) from the portions of the detectors
not covered by the long slit, and subtracted it from the scientific
spectra. This ensures that the mean contribution from any scattered
source is removed. The error in the quantification of the scattered
light is very similar to the error in the sky subtraction, which does
not explain the observed difference in stellar population properties
(Fig. \ref{fig:ssp}). Small residual variations of the scattered light
intensity along the long-slit might still be present, but it is not
possible to measure them. Nevertheless, because the light is scattered,
we expect that it contaminates adjacent regions of the detector by a
similar amount. On the other hand, we observe a large difference between
the spectral properties of the adjacent bins D and E, suggesting that
this difference reflects a variation in the stellar population
properties in different slit sections.

\section{Discussion}
\label{sec:discussion}

NGC 3311 is surrounded by a dynamically hot stellar halo,
characterized by a rapid increase of the velocity dispersion of the
stars \citep{Hau+04, Loubser+08, Ventimiglia+10} and of the globular
clusters system \citep{Richtler+11}. In this section we will use the
information on the stellar population derived in Sect. \ref{sec:ssp}
to constrain the origin of this dynamically hot stellar halo.

In this region, \citet{Arnaboldi+11} identified a photometric
sub-structure (excess of light), which extends toward the northern
and eastern sides of NGC 3311. In Fig. \ref{fig:slit_location} we show
the approximate contours of the excess of light in the field of
view. An accurate determination of the morphology, photometry and
precise isophotes of the excess is complicated by the presence of
foreground stars and uncertainties in the background subtraction and
 is not the aim of the present paper. For our purposes it is enough
to notice that the FORS2 long-slit crosses the excess of light from SE
to NW, probing regions where its intensity is different. Although the
isophotes in Fig. \ref{fig:slit_location} are approximate, we expect
that the contribution from the excess of light to the mean stellar
population properties along the line of sight has a minimum in bin A
and a maximum in bin E. From Fig. \ref{fig:ssp} we have evidence that
the mean stellar populations in bins E and D have different
metallicities from those observed in the other slit sections. We
interpret this difference as caused by the contamination from the excess
of light, which is higher in bin E (where the difference in
metallicity is more evident) and lower in bin D (where it is less
evident).

We therefore split the discussion into two parts: Section
\ref{sec:halo} describes the stellar population properties inferred
from the portion of the slit covered by bins A-C, while Section
\ref{sec:excess} describes those derived from bins D-E.

\subsection{The dynamically hot stellar halo around NGC 3311}
\label{sec:halo}

The spectra extracted in regions A-C contain the contribution mostly
from stars in the dynamically hot halo around NGC~3311. Indeed, the
rapid increase over radius of the velocity dispersion (see Section
\ref{sec:discussion}) suggests that the contribution of stars bound to
the main body of NGC~3311 at these galactocentric distances is
small. From the contour levels displayed in
Fig. \ref{fig:slit_location} we can conclude that the contribution
from the excess of light is also small.  Therefore, we consider the
age, [Z/H] and [$\alpha$/Fe] measured in bins A-C as indicative of the
mean luminosity-weighted stellar population properties in the
dynamically hot halo around NGC 3311, dominated by intra-cluster
stars.

In this region, stars are old (age $>13$ Gyr), metal-poor ([Z/H]$=
-0.35 \pm 0.05 $) and formed over a rapid time scale ($<1$ Gyr),
as inferred from the high abundance ratio of $\alpha$ elements
([$\alpha$/Fe] $=0.48 \pm 0.03$). Similar values are measured in the
outer regions of other BCGs, e.g. NGC 3557 \citep{Brough+07}, NGC 1400
\citep{Spolaor+08}, NGC 1407 \citep{Spolaor+08, Forbes+11}. But,
contrarily to NGC 3311, the velocity dispersion profiles in the halos
of these galaxies are either decreasing or nearly constant, indicating
that the stars are still bound to the galaxy and that the
contribution from intra-cluster stars to the kinematics is small.  The
stellar halo around NGC 3311, however, consists of mostly intracluster
stars. The lack of a significant metallicity gradient over the spatial
range probed by slit sections A-C indicates that the stars
accumulated on the halo through several accretion episodes. What are
the progenitors of these stars?

\smallskip 

\noindent 1.  Stars from the central regions of massive early
type galaxies are most likely ruled out.  The distribution of
their [Z/H] ([$\alpha$/Fe]) has mean 0.35 (0.19) and dispersion 0.30
(0.07) in solar units. These average values are based on measurements in
early-type galaxies in the Coma cluster with velocity dispersion
$\sigma > 200$ \kms, from \citet{Trager+08}.  The stellar population
properties measured in the central regions of the early-type galaxies
in Hydra I where measurements of the Lick indices are available (IC
2586, IC 2597, NGC 3308, ESO436-G04, and ESO501-G13)\footnote{See
  \citealt{Jorgensen97, Trager+98, Ogando+08} for the line strength of
  the Lick indices.\label{foot}} are consistent with the average
values in Coma.

\smallskip 

\noindent 2. Stars in early-type galaxies outside 1-2 effective radii
are potential candidates. Indeed, several early-type galaxies in the
sample of \citet{Spolaor+10}, \citet{Pu+10} have at 1-2 effective
radii [Z/H] and [$\alpha$/Fe] similar to what we observe in the halo
of NGC 3311.  Unfortunately, no data outside of one effective radius are
available for early-type galaxies in Hydra I, but only central values
for a number of them (NGC 3305, NGC 3309, ESO436-G45 and
ESO501-G13)$^{\ref{foot}}$. They are characterized by alpha enhanced
stars ([$\alpha$/Fe]$>0.4$) at their centers, but they have
[Z/H]$>0$. Stars stripped from their halos might be consistent with
the stellar population of the stars around NGC 3311 under the
(plausible) assumption that these galaxies are characterized by strong
and negative metallicity gradients, and flat [$\alpha$/Fe] radial
profiles.

\smallskip 

\noindent 3. Stars stripped from the disks or the bulges of late-type
spirals are most likely ruled out. They are generally characterized by
younger ages and by solar values of metallicity and $\alpha$-elements
abundance \citep{Crowl+08, Morelli+08}.  The stellar population
properties measured in the central regions of the late-type galaxies
in Hydra I where measurements of the Lick indices are available
(ESO437-G015, ESO437-G45, and ESO501-G03)$^{\ref{foot}}$ are consistent
with these values. A small fraction in mass of young stars would
decrease the luminosity-weighted age of the corresponding single
stellar population \citep{Serra+07}, in contrast with our
measurements.

\smallskip

\noindent 4. Stars from disrupted dwarf galaxies are potential
candidates. They are characterized on average by subsolar metallicity
and nearly solar $\alpha$-enhancement; e.g., in the sample of
\citet{Michielsen+08}, the distribution of [Z/H] ([$\alpha$/Fe]) in
dwarfs has mean value $-0.4$ (0.0) with a scatter of $0.3$ (0.2) dex.
Therefore, stars from disrupted dwarf galaxies can lower the
luminosity weighted metallicity of the equivalent single stellar
population. On the other hand, their contribution would also decrease
the luminosity-weighted [$\alpha$/Fe], meaning that they cannot
represent the main bulk of stars in the NGC 3311 halo.

\smallskip 

In conclusion, the combination of kinematic and stellar population
data suggests that the hot stellar halo around NGC 3311 was accreted.
The bulk of the stars most likely originated from the outer regions of
early-type galaxies with metal poor and $\alpha$-enhanced stellar
populations. Stars from disrupted dwarf galaxies are also potential
contributors, because of their generally low [Z/H].

\subsection{The origin of the excess of light}
\label{sec:excess}

As discussed above, we have evidence that the stellar population
measured in the long-slit bins D-E is different from that measured in
bins A-C.  Following the same dynamical and photometric arguments as in
Sect. \ref{sec:halo}, we consider the contribution of stars bound to
NGC 3311 to be small, but not the contribution of the excess of
light. Therefore, there are potentially two main stellar populations
in this region, and the approximation of single stellar population
adopted in Sect.  \ref{sec:ssp} may not be the appropriate way to
interpret the observations.

Despite of the $S/N \geq 25$ (per pixel) reached and the good quality
of the spectra, we have been unable to separate the two populations by
simultaneously fitting their kinematics and stellar populations
(e.g., using a direct spectral fit as in \citealt{Coccato+11}). This is
probably because of the high velocity dispersion that broadens and
mixes the spectral features of the two components along the line of
sight.

A better way to interpret our observations is therefore to make some
assumptions on the kinematics of the excess of light and on the nature
of its progenitors on the basis of available additional
information. Following these assumptions, we created a simulated
spectrum $S_{\rm model}$ from the combined contribution of the stellar
halo and the excess of light and tested whether it is consistent with
what is observed.
 
\citet{Arnaboldi+11} were able to estimate the kinematics of the
excess of light from the cross correlation between a template star and
a model spectrum for the excess. They found two positive peaks on the
correlation function above $\sim 2\sigma$ noise. The first peak was at
5032 \kms, and it was interpreted as the velocity of the excess of
light along the line of sight. The second was at 3905 \kms, consistent
with the systemic velocity of NGC 3311 and it was interpreted as the
residual contamination of the stars from the halo in the model
spectrum of the excess of light. Therefore the relative velocity of
the excess of light with respect to NGC 3311 is on the order of 1200
\kms. \citet{Arnaboldi+11} also propose a scenario in which the excess
of light originates from disrupted dwarf galaxies and they estimate
that the excess of light represents 10\% -- 20\% of the surface
brightness of the halo of NGC~3311 in the region where the excess is
detected.

Following their interpretation, we assume that we observe the
superimposition of the halo stellar population and a dwarf-galaxy-like
population in bins D-E. As template for the halo spectra, we used that
extracted from bin A; as template for a dwarf galaxy spectrum we used
that of the dwarf galaxy HCC~26, on which the FORS2 long-slit was
centered (see the orange circle in Fig. \ref{fig:slit_location}). The
stellar population properties of HCC26 are consistent with those of
other dwarf galaxies \citep{Michielsen+08}. As a first approximation,
we normalized the spectra of bin A and HCC~26 to their continuum
values at 5100 \AA. We indicate these normalized spectra with $S_{\rm
  halo}$ and $S_{\rm dwarf}$, respectively. We then convolved $S_{\rm
  dwarf}$ with a Gaussian LOSVD to construct the contaminant component
$\tilde S_{\rm dwarf}$. The LOSVD we used for the convolution has mean
velocity 1200 \kms\ higher than the systemic velocity of NGC 3311 and
velocity dispersion of 450 \kms. These assumptions are consistent with
the estimate of the velocity of the excess of light with respect the
halo around NGC 3311 \citep{Arnaboldi+11}, the typical velocity
dispersion measured in these regions \citep{Ventimiglia+10} and the
standard deviation of the line-of-sight velocities of the seven dwarf
galaxies located on the top of the excess of light ($\sigma_{\rm
  7\ DWARFS} \sim 470$ \kms, \citealt{Misgeld+08}), which are
potential progenitors.\footnote{The results do not vary too much over
  the velocity dispersion range 200 \kms $\lesssim \sigma \lesssim$
  600 \kms, meaning that we cannot infer precise constraints on the
  kinematics using the stellar population properties.} We then
constructed many combined spectra $S_{\rm model} = (1-F) \cdot S_{\rm
  halo} + F \cdot \tilde S_{\rm dwarf}$ for different values of the
``contaminating'' fraction $F$, measured their equivalent single
stellar population values of age, [Z/H] and [$\alpha$/Fe], and compared
them with the observed stellar population properties in bins D and
E. We prefer this strategy over a direct comparison of the observed
and combined spectra $S_{\rm model}$ because it is less sensitive to
the uncertainties of the overall shape of the stellar continuum. Lick
indices are less affected by these uncertainties, because the
pseudo-continua and the pseudo-bands where each index is defined are
very close in wavelength.

\begin{figure}
  \psfig{file=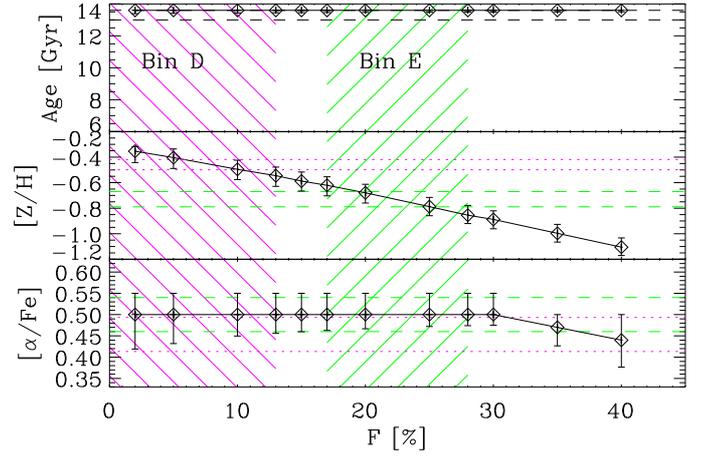,width=9.0cm,clip=}
\caption{Equivalent single stellar population age (upper panel),
  metallicity (central panel) and $\alpha$-enhancement (lower panel)
  measured in $S_{\rm model}$.  $S_{\rm model}$ represent a composite
  stellar population formed by stars from the halo around NGC~3311
  contaminated by a fraction $F$ of stars from dwarf galaxies (see
  Sect. \ref{sec:excess} for details). Long-dashed green and
  short-dashed magenta lines represent the mean values of age, [Z/H]
  and [$\alpha$/Fe] measured in the long-slit bins E and D,
  respectively. The range of $F$ values consistent with measurements in
  bins E and D are shaded in green and magenta, respectively.}
\label{fig:contamination}
\end{figure}

Figure \ref{fig:contamination} shows the results as function of
$F$. The mean values of the single stellar population parameters
measured in bins D-E are consistent with those of a composite
population made of stars from the stellar halo around NGC 3311 and
stars from disrupted dwarf galaxies. In bin E the contamination is
stronger than in bin D. In detail, the range of contamination in the
surface brightness from 17\% to 28\% reproduces within uncertainties
the mean values of age, [Z/H] and [$\alpha$/Fe] measured in bin E
(between horizontal green long-dashed lines). A smaller fraction ($<$
13\%) is needed to explain the mean values observed in bin D (between
horizontal magenta short-dashed lines). The ranges of contamination in
surface brightness we found are on the same order of magnitude of the
estimate in \citet{Arnaboldi+11}.

The presence of a contaminant component with lower velocity dispersion
($\sigma \lesssim 60$ \kms) was also tested, and we found that it can
contribute less than $F\leq$5\% to the local surface brightness in all
bins. Higher fractions would produce absorption lines features clearly
visible on the combined spectrum (contrarily to what we observe in
bins D-E), which would be recovered by a direct spectral fit
\citep{Coccato+11}.

Despite of the uncertainties on the kinematics that we assumed for the
contaminant stellar component, this simple approach supports the idea
that the excess of light in the Hydra I cluster core consists of stars
stripped from dwarf galaxies, which are thus being added to the
dynamically hot stellar halo around NGC 3311.

\section{Summary}
\label{sec:summary}

We used deep long-slit spectra obtained with FORS2 at the VLT on the
northeast side of the brightest cluster galaxy NGC 3311 to study the
stellar population content in five regions from $\sim$10 to $\sim$30 kpc
from its center (slit sections A-E), where the contribution from
intra-cluster stars dominates \citep{Ventimiglia+10}.

We found that the stars of the dynamically hot halo around NGC 3311
are on average characterized by an old, metal-poor and
$\alpha$-enhanced stellar population. The lack of a stellar population
gradient in slit sections A-C together with the kinematics suggest
that the intracluster halo formed from accretion of multiple
components onto the cluster center. The stellar population parameters
suggest that the halo stars were mainly stars stripped from the outer
regions of early-type galaxies, with a possible contribution from
dwarf galaxies.

In the long-slit sections D-E the metallicity is significantly lower
than that measured in sections A-C. We associate this spectroscopic
feature with the presence of a photometric substructure (excess of
light), which was detected in the same area of the halo around NGC
3311 \citep{Arnaboldi+11}. Simulations of the measurements over this
region are consistent with the presence of a composite population made
of stars from the stellar halo around NGC 3311 and stars from
disrupted dwarf galaxies, which form part of the excess of light. Our
findings therefore support the idea that the build-up of ICL in the
Hydra I cluster core and the stellar halo around NGC~3311 are still
ongoing.

\begin{acknowledgements}
The authors wish to thank H. Kuntschner and L. Morelli for useful
discussions on the measurement of the Lick indices.
\end{acknowledgements}
\bibliography{coccato11}
\end{document}